\documentclass[a4paper,11pt]{article}
\usepackage{jinstpub} % for details on the use of the package, please see the JINST-author-manual
\usepackage{lineno}
\usepackage{tablefootnote}
\title{\boldmath Extending the time-over-threshold calibration of Timepix3 for spatial-resolved ion spectroscopy}

 \author[a,b,1]{R.E. Mihai,\note{Corresponding author.}}
 \author[a]{B. Bergmann,}
 \author[a]{and P. Smolyanskiy}
 \affiliation[a]{Institute of Experimental and Applied Physics, Czech Technical University in Prague,\\Husova 240/5 110 00 Prague 1, Czech Republic}
 \affiliation[b]{Horia Hulubei National Institute for R\&D in Physics and Nuclear Engineering,\\No. 30 Reactorului Street, M\u agurele, Ilfov, Romania}

\emailAdd{mihairad@cvut.cz}

\abstract{Hybrid pixel detectors have a well-established array of applications ranging from particle physics to life sciences. The small dimensions of Timepix3 as well as its relatively low energetic expenses make it an intriguing option additionally for ion detection in nuclear physics experiments, as it reveals simultaneously precise temporal, spatial and energetic properties of recorded events from nuclear reactions. Currently, a limiting factor is the electronics behavior at high input charge resulting in improper energy determination of incident heavier ions. While the low-energy per-pixel calibration of Timepix3 is normally performed with the use of photons of up to 60 keV, the characteristic linear range permits a correct extrapolation up to only 150 keV/pixel. We developed a global per-pixel energy correction method involving the use of short-ranged accelerated ions and spectroscopic alpha sources, to suitably extend the energy determination capability of Timepix3 for nuclear ion spectroscopy experiments, where spatial and temporal precision of recorded events are equally crucial. It was found that upon applying this correction, the reliable per-pixel energy range has been increased from the original 150 keV to at least 1.1 MeV, while maintaining the relative energy resolution to better than 2.5\% for stopped protons of up to 1.9 MeV and better than 3.1\% for $\alpha$-particles of 5.5 MeV. Furthermore, to demonstrate the spatial resolution of Timepix3 detectors with silicon sensors, we present $\alpha$-radiography measurements from which we extract the modulation transfer function (MTF) and produce real-world biological sample images. }

\arxivnumber{2501.04419} % Only if you have one

\begin{document}
\maketitle
\flushbottom

\section{Introduction}
\label{sec:intro}
Timepix-family hybrid pixel detectors are used in a variety of fields ranging from radiation environment monitoring, space radiation dosimetry, and space weather studies to life sciences~\cite{Ballabriga,Jakubek_life,Bergmann_MDPI}. Most of these applications profit from the high spatial granularity of the detectors, providing radiation imaging capabilities or particle species discrimination with a single-layer device of small dimensions and at relatively low energetic expense.  

In comparison with Timepix~\cite{Llopart2007} and Timepix2~\cite{Wong2020}, a particular advantage of Timepix3~\cite{tpx3general} is the combination of nanosecond-scale time resolution, simultaneous per-pixel time and energy determination, with dead-time free measurement. This provides means for 3D reconstruction of particle trajectories~\cite{Bergmann_3D, Bergmann_3D_CdTe}, allows for precise measurements of radioisotope half-life times of the order of nanoseconds~\cite{Bergmann_PO,bergmann_seminar}, as well as time-of-flight determination for assessment of particle energy~\cite{tpx3neutrons}. 

The current limitation of Timepix3, is a deteriorated spectroscopic response at high per pixel input charge. This prevents its proper use in nuclear physics experiments, where a measurement of the energy of ions fully absorbed in the sensor is crucial, or in high energy heavy ion beams, where proper determination of dE/dX is needed for identifying the nuclear charge $Z$ of the ions (see e.g.,~\cite{tpx3saturation}). While for Timepix and Timepix2 the per-pixel energy measurement and its dependence on incident energy has been studied in previous works~\cite{tpx2adaptive,tot_clb}, an equivalent study for Timepix3 is missing. Up to date, a similar response as for Timepix, discussed for example in~\cite{George2018}, is expected. This means a linear response at low energies is followed by a saturation level for intermediate energy deposition, while towards the highest energy deposition, a decrease with increasing energy is observed. For Timepix3, the saturation level was reported to be at 500\,keV~\cite{Kroupa2016,tpx3saturation}.

The key challenges of developing energy calibration methodology beyond $\sim$60\,keV per-pixel energy are that particles must deposit all of their energy within the volume of a single pixel, while the locally created charge cloud of the interaction must not spread out across multiple pixels \cite{tot_clb}. 
While the former is the key issue for calibration with higher energy $\gamma$-rays or electrons, the latter is a particular problem when intending to calibrate with short-ranged protons or $\alpha$-particles.

The present work addresses these issues and presents an energy correction method involving the use of short-ranged accelerated ions and spectroscopic $\alpha$-sources, similar to the one described in \cite{tpx2adaptive}. We present the energy spectra after applying the correction in the range from $\sim$500\,keV to 5.5\,MeV and we show, as an example of practical applications, $\alpha$-particle radiography measurements with different samples.

\section{Experiment}
\subsection{Timepix3}
Timepix3~\cite{tpx3general} detectors consist of the readout ASIC developed within the Medipix3 collaboration, which is connected to a sensor. Both ASIC and sensor are segmented into a square matrix of 256~$\times$~256 pixels of 55 $\times$ 55 \,\textmu m$^{2}$ dimensions. In each pixel, the energy and the timestamp of a particle interaction are measured simultaneously, the latter being achieved with a precision of $\sim$2\,ns. 

Ionizing particles interacting in the silicon sensor create charge carriers, which drift along the electric field lines created through reverse biasing towards their corresponding electrode. In the silicon sensors used, electrons drift towards the pixelated cathode while the holes drift towards the common backside anode. During this drift motion, the charge carriers induce signals at the pixel electrodes closest to their location, which are amplified and shaped in a charge-sensitive preamplifier in the pixel's electronic circuits. The charge-sensitive amplifier (CSA) output pulse is then compared to a globally adjustable threshold (THL). A 40\,MHz continuous running clock is  used to measure the duration in which the pulse is above THL (ToT), while the moment when the pulse crosses THL defines the time of arrival (ToA). In order to improve ToA precision, an additional 640\,MHz clock from local oscillators is used to sample the time from the actual THL crossing until the next rising edge of the base clock. While noise-free operation is possible at THL of around 3\,keV for the study in the present work, THL has been set conservatively to 5\,keV.

Within the scope of the present work, Timepix3 was exclusively used in the so-called data-driven mode, in which data are sent off chip on a pixel-by-pixel base where only pixels which were triggered by an interaction are read out while all other pixels remain active. The pixel readout (per-pixel dead) time is 475\,ns. The resulting stream of hit pixel data is then used as input for the event building, which groups spatially and temporally neighboring pixels into "clusters" which are assumed to correspond to individual particles interacting in the sensor.

\subsection{Experimental setup}
\begin{figure}[htbp]
\centering
\includegraphics[width=.49\textwidth]{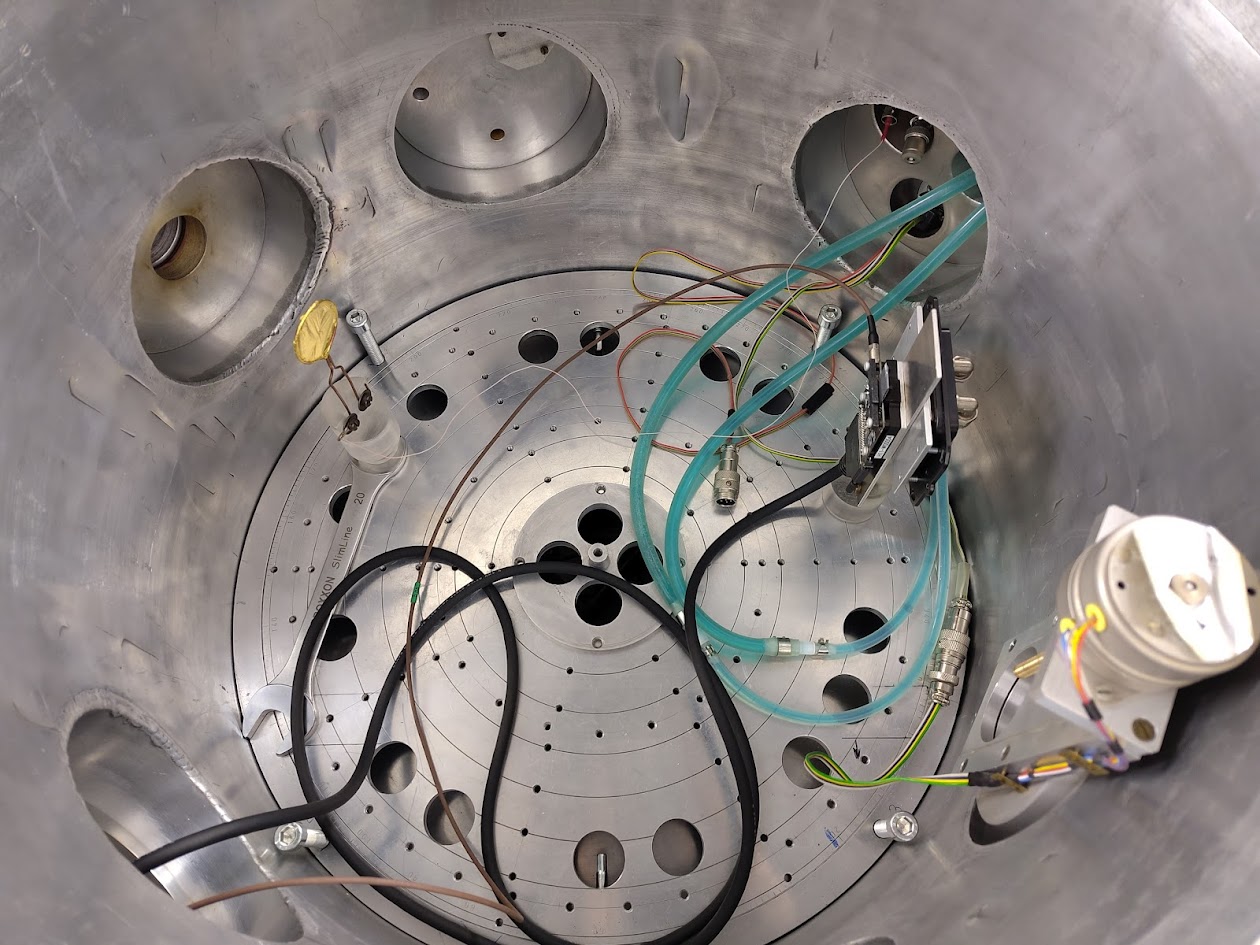}
\includegraphics[width=.49\textwidth]{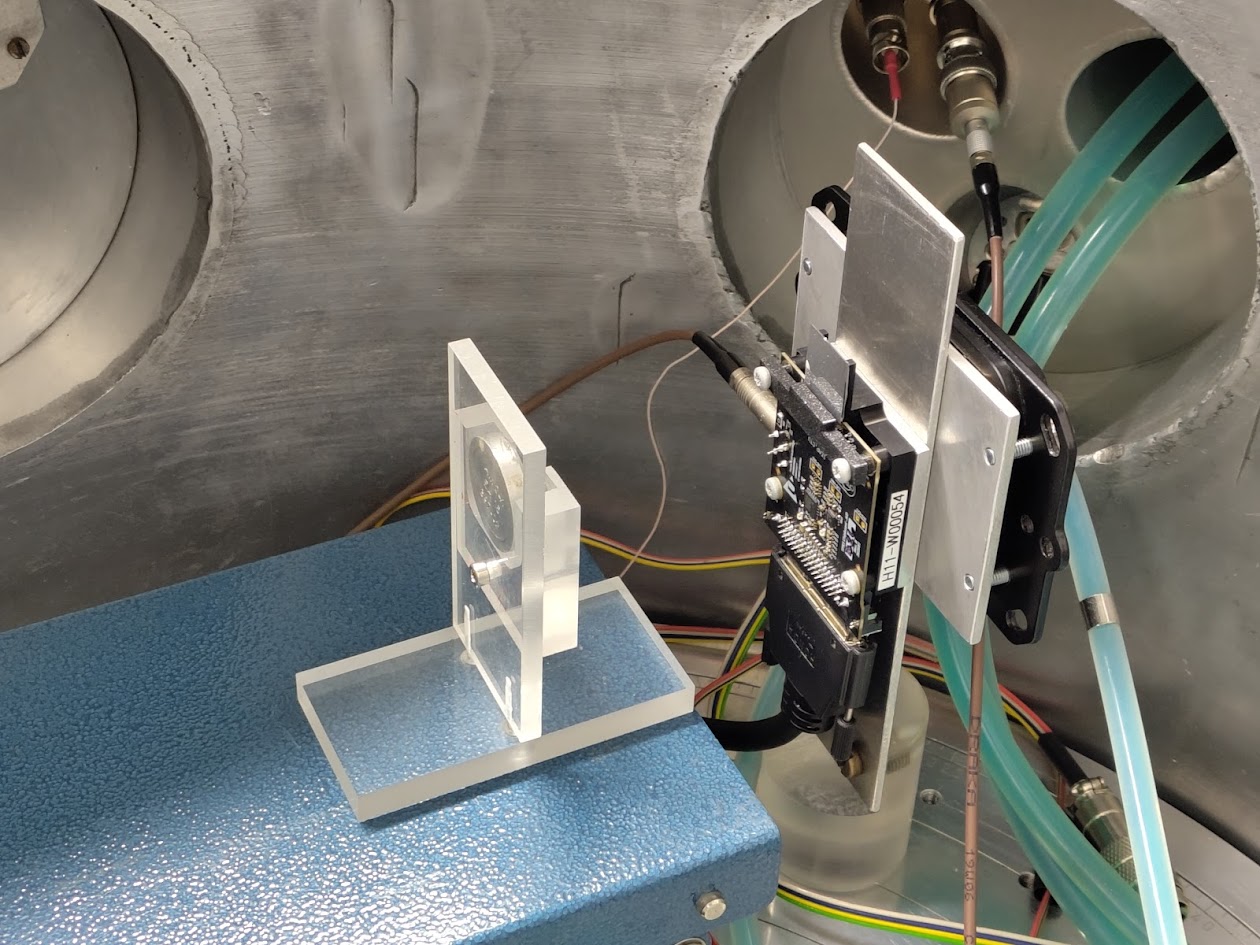}
\caption{In-vacuum detection setups for the measurement of RBS protons(\textit{left}) and $^{241}$Am alpha particles(\textit{right}) with Timepix3 detectors.
\label{fig:setups}}
\end{figure}

The proposed method required two types of measurements with silicon-based Timepix3 detectors: firstly, mono-energetic protons in the range from 500\,keV to 1.9\,MeV, delivered by an in-house Van-de-Graaff accelerator~\cite{webpage}, were shot at a gold foil of 100-nm thickness. The detectors were placed at 135$^\circ$ with respect to the beam axis to measure back-scattered protons (see Fig.~\ref{fig:setups}). In a second experiment, $\alpha$-particles emitted by a $^{241}$Am source were measured with different bias voltage settings - ranging from 100\,V to 150\,V for the 300\,\textmu m sensor and from 200\,V to 300\,V for the 500\,\textmu m sensor. Bias voltage variation allows to study at different maximal per-pixel recorded energy, due to charge expansion during the drift towards pixel electrodes. The voltage range was selected to satisfy the condition of complete sensor depletion and to also prevent excessive degradation due to high-current-induced electro-migration. All mentioned measurements were performed in vacuum with the aid of liquid cooling for thermal stability.

After the collision between scattered protons and the thin gold target, the former conserves its kinetic energy directly proportional to the square of a kinematic factor $k$ of the following form: 
\begin{equation}
\label{eq:x}
\begin{aligned}
k &= \frac{m_p \cdot \cos{\theta}+\sqrt{m_{Au}^2-m_p^2 \cdot \sin^2{\theta}}}{m_p+m_{Au}},
\end{aligned}
\end{equation}
\noindent where $m_p$ and $m_{Au}$ are the masses of protons and gold atoms, respectively, while $\theta$ is the scattering angle. Therefore, the relation between the energy before ($E_{p_i}$) and after back-scattering ($E_{p_f}$) is
\begin{equation}
\label{eq:y}
\begin{aligned}
E_{p_f} &= k^2 \cdot E_{p_i}  \,.\\
\end{aligned}
\end{equation}

Incident particle energies were determined through SRIM simulations \cite{f} assuming metalized backside contacts of 0.5 \textmu m- and 1 \textmu m-thick aluminum, and doped-region dead layers of 0.8 \textmu m- and 2.5 \textmu m-thick layers of silicon for the 300-\textmu m and 500-\textmu m sensors, respectively.

\newpage
\begin{table}[htbp]
\centering
\caption{Properties of incident particles used in the calibration procedure for a Timepix3 detector with a 500-\textmu m silicon sensor. \label{tab:j}}
\smallskip
\begin{tabular}{c|c|c|cc}
\hline
\hline Particle & Energy at emission & Energy after RBS & Energy after dead layer \\
 & (keV) & (keV)& (keV) \\ 
\hline
p & 600 & 588 & 392   \\
 & 700 & 685 & 512   \\
 & 800 & 784 & 627   \\
 & 900 & 882 & 739  \\
 & 1000 & 980 & 860   \\
 & 1250 & 1224 & 1118   \\
 & 1500 & 1469 & 1382   \\
 & 1750 & 1714 & 1642   \\
 & 1900 & 1861 & 1797   \\
 \hline
$\alpha$ & 5468\tablefootnote{$\alpha$-particle energy is an intensity-weighted average of $^{241}$Am $\alpha$ emissions, due to their likeness being below the resolving power of silicon sensors.} & - & 4930 \\
\hline
\end{tabular}
\end{table}

\section{Methodology}

\begin{figure}[htbp]
\centering
\includegraphics[width=.49\textwidth,trim={0 0 10cm 0},clip]{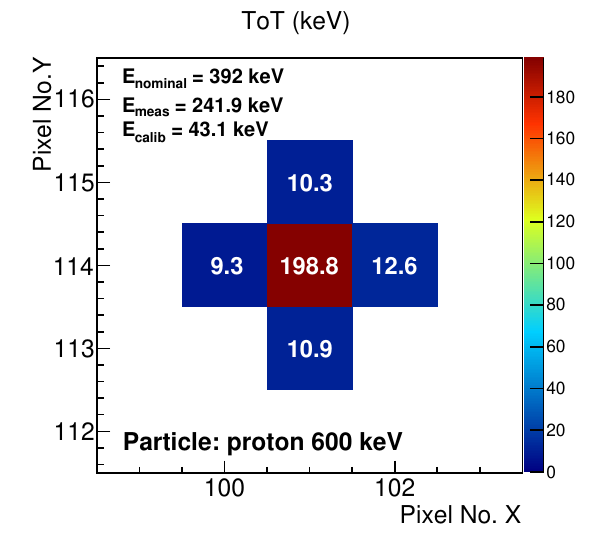}
\includegraphics[width=.49\textwidth,trim={0 0 10cm 0},clip]{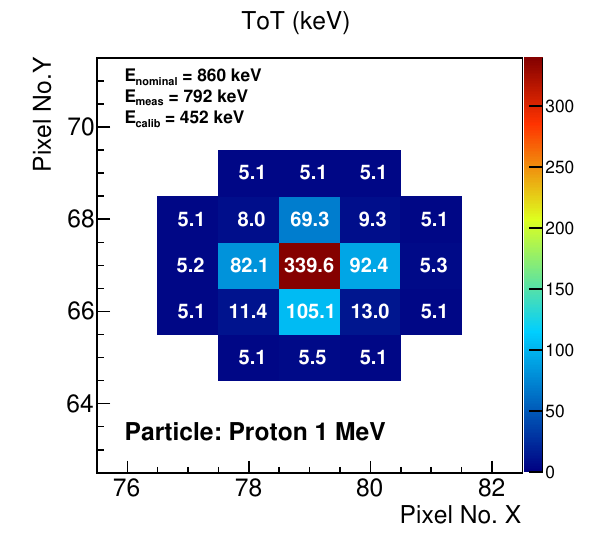}
\caption{Example events used for determination of the high energy pixel correction. (\textit{left}) A cluster created by a 600\,keV proton; (\textit{right}) cluster created by a 1 MeV proton. In both cases only the central pixel is in the "uncalibrated" energy region.
 }
\label{fig:illustration}
\end{figure}

The procedure starts by assuming the energy calibration performed with X-rays is valid up to a per-pixel energy value of 150 keV. The pixels within a cluster are then put into the two categories, "calibrated" (pixel of energy below 150\,keV) and "uncalibrated" (others). Example clusters for the measurements at 600\,keV and 1\,MeV proton impact are shown in Fig.~\ref{fig:illustration}. In both cases, only the central pixel (red) is "uncalibrated" while the measured energy $E_{\textrm{meas}}$ is lower than the nominal energy $E_{\textrm{nominal}}$.

Defining the sum of the energies of the set of calibrated pixels $E_{\textrm{calib}}$, the difference $E_{\textrm{nominal}}- E_{\textrm{calib}}$ describes the energy needed in the central pixel $E_{\textrm{expected}}$ for proper energy measurement. Thus, by evaluating the dependence of $E_{\textrm{expected}}$ as a function of the energy measured in the uncalibrated pixel $E_{\textrm{meas}}$, the correction function is determined. If the set of uncalibrated pixels has more than one and less than 5 pixels with a maximal difference of energies below 5\%, the missing energy is equally distributed among the uncalibrated pixels, the expected energy is given by $E_{\textrm{expected}} = \frac{E_{\textrm{nominal}}- E_{\textrm{calib}}}{N_{\textrm{uncalibrated}}}$. 

Since the number of pixels and the dispersion of pixel energy increases with increasing energy deposition, the $\alpha$-particle measurements could not be used with the 150-keV threshold for pixel categorization. Therefore, the calibration procedure is done in two steps. First, the correction function is determined for the proton data sets (iteration 1). Then the likewise determined calibration is applied and the method is applied to $\alpha$-particle data but with a raised threshold (600\,keV) for categorization as uncalibrated (iteration 2). 

\begin{figure}[htbp]
\centering
\includegraphics[width=.49\textwidth]{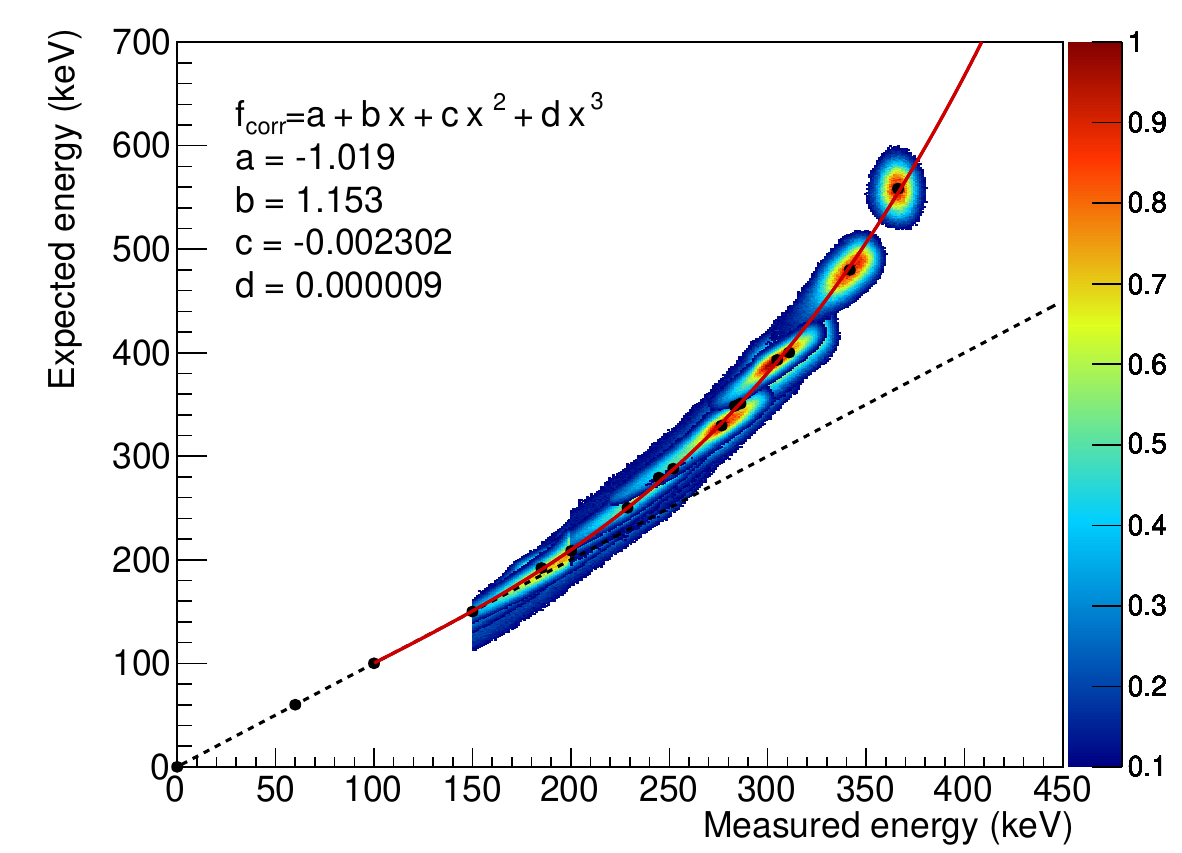}
\includegraphics[width=.49\textwidth]{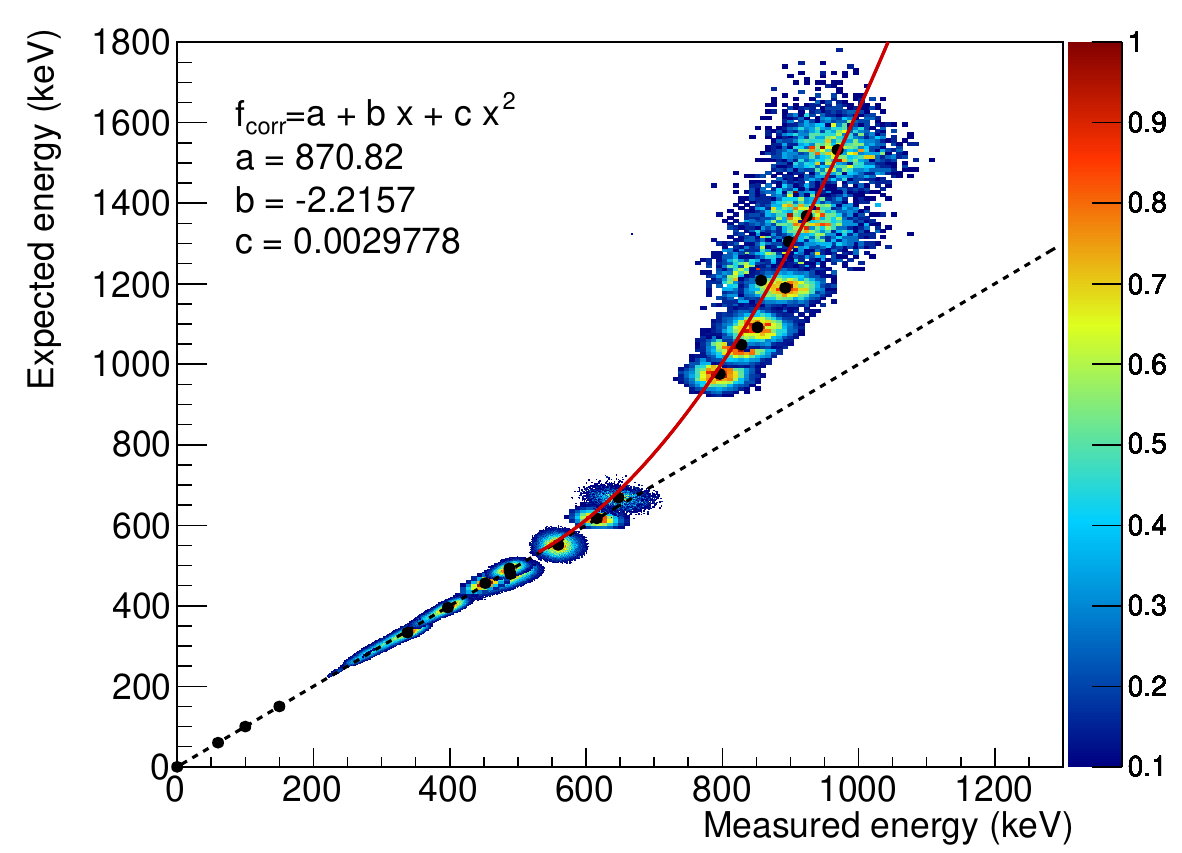}
\caption{Iterative global calibration curves(red) for events registered by individual Timepix3 chip pixels connected to a 500-\,\textmu m silicon sensor. For the first iteration (\textit{left}) - expected energies up to 600 keV were taken into account and corrected with a third degree polynomial fit, while for the second iteration (\textit{right}), events up to 1.6 MeV expected energies were included and corrected with a second degree polynomial fit. Protons in the 0.5 -- 1.9 MeV energy range and $\alpha$-particles of 5.5 MeV measured at different bias voltages were used for the calibration curves depicted in the left and right graph, respectively.
\label{fig:iterations}}
\end{figure}

The results of iteration~1 are shown in Fig.~\ref{fig:iterations}~(left). The correction function was fitted with a third order polynomial. The results of iteration~2 are shown in Fig.~\ref{fig:iterations}~(right). The per-pixel energy correction function of iteration~1 provides proper energy measurement up to at least 600\,keV, while significant deviations from linearity become visible at 800\,keV. The subsequent deviations were less complex, therefore the correction function for the second iteration only required the use of a second-order polynomial.

For conciseness, the calibration methodology is demonstrated for a 500\,\textmu m in the main body of the manuscript, the equivalent figures for a 300\,\textmu m thick sensor are shown in Annex A. The correction fits of the first iteration were observed to be applicable on data taken with all tested Timepix3 silicon detectors, leading to less than 2\% measured cluster energy difference between devices for protons up to 1.75 MeV, regardless of sensor thickness, after correcting for specific dead layer energy losses. The second iteration of the correction is more sensor-dependent, the values diverging from a unitary ratio at different points due to proprietary characteristics of silicon sensors at different bias voltages. It is therefore trivial to conclude that, in a first approximation, the first set of calibration curve coefficients presented in the current work can be considered universal for Timepix3 silicon detectors, provided that the classical X-ray calibration is correctly performed and applied beforehand, while the second set differs from one sensor to another and should be determined individually. Moreover, data acquired in different $I_{\textrm{krum}}$, energy threshold settings and running modes with several sensors revealed no real necessity of altering the first iteration correction parameters, provided that the appropriate initial X-ray calibration is applied.

\section{Results}
\subsection{Energy resolution}
\begin{figure}[htbp]
\centering
\includegraphics[width=.89\textwidth]{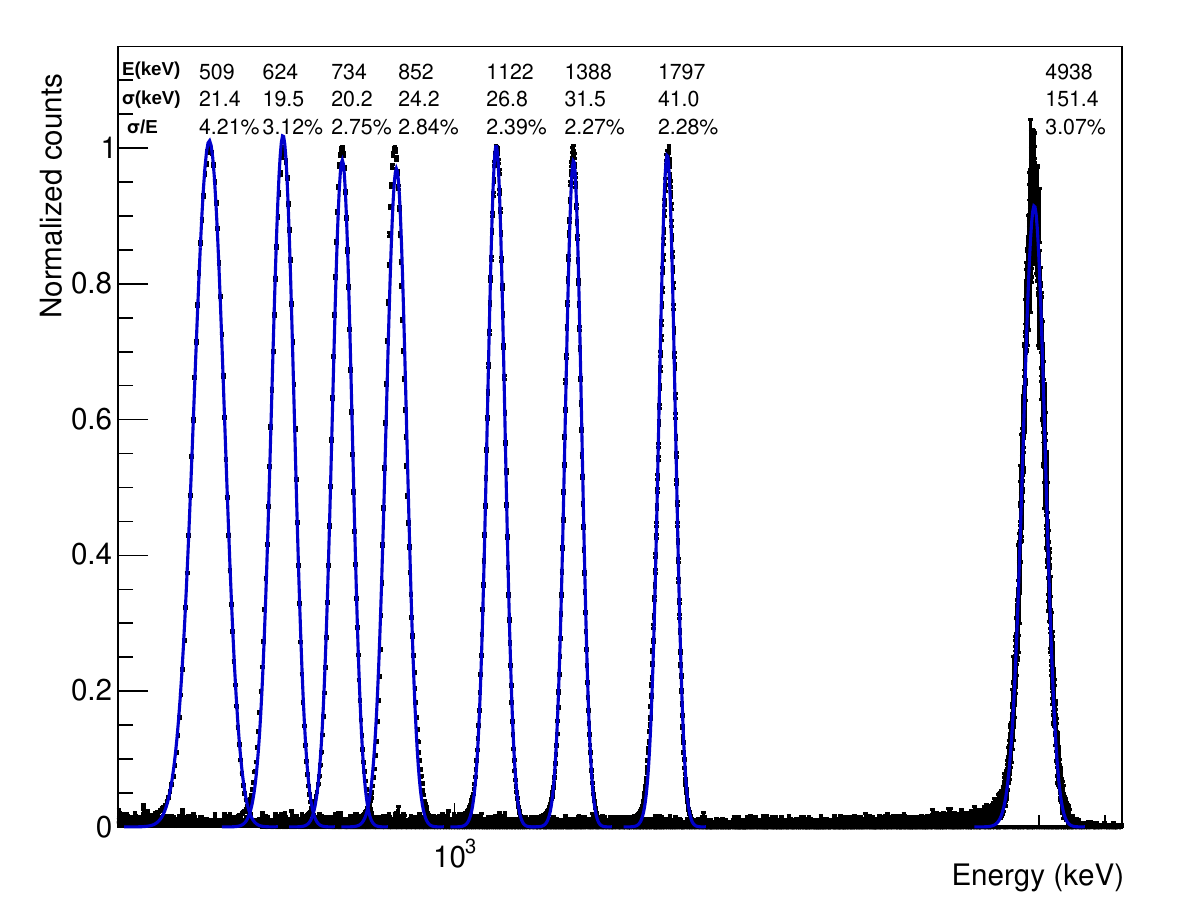}
\qquad
\caption{Energy spectra of experimentally measured incident protons and alpha particles with a 500 \textmu m Si Timepix3 detector after applying the iterative correction method described in the present work. Intensities are scaled to the lowest height.
\label{fig:energy_calibration}}
\end{figure}

Figure~\ref{fig:energy_calibration} shows the corrected energy spectra for proton and $\alpha$-particle impact of 700\,keV up to 1.9\,MeV and 5.5\,MeV, respectively. As discussed above, due to particle energy losses in the dead-layer, the overall measured energy is lower than the incident energy. Overall,  a relative energy resolution of around and below 3\% was achieved.

\subsection{$\alpha$-particle radiography}
\subsubsection{Slanted edge measurement - spatial resolution}
The modulation transfer function (MTF) is a means of assessing the imaging performance of optical systems by determining the level of achievable contrast. One of the most common methods of determining the MTF is the slanted edge imaging technique (certified under ISO12233), through which spatial frequency information can be obtained by placing a sample, with homogeneous properties and a straight edge, in the imaging device's field of view at a relatively small angle, partially covering individual pixels. In this case, the irradiance distribution is an edge spread function (ESF). The derivative of the ESF is then employed to obtain the line spread function (LSF), and its Fourier transform will result in the MTF itself \cite{e}.

\begin{figure}[htbp]
\centering
\includegraphics[width=.99\textwidth]{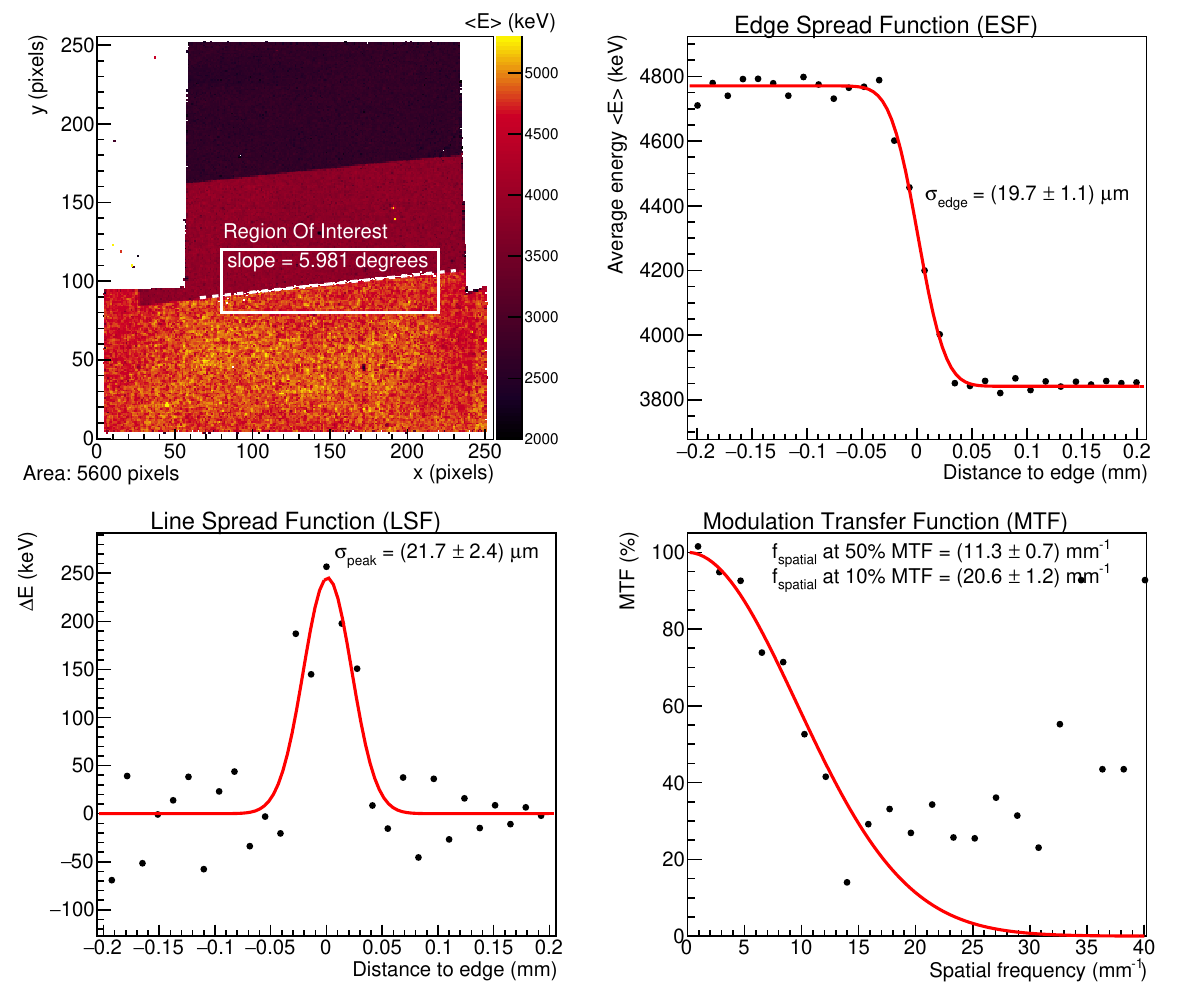}
%\includegraphics[trim={10cm 0 0 0}, width=.49\textwidth, clip]{eneergy_resolution_slanted_edge.pdf}
%\qquad
\caption{Application of the slanted edge method for determining the spatial resolution in Timepix3 $\alpha$-particle radiography: (a) 2D map of the average energy measurement across the pixel matrix. The Region of interest used for determination of the MTF is indicated; (b) Edge spread function including edge fit determined from the ROI in (a); (c) Line spread function given as the derivative of the ESF in (b). It is modeled with a Gaussian; (d) Modulation Transfer Function (MTF) describing the spatial resolution of the imaging system. It was determined as the Fourier Transform of the LSF in (c).}
\label{fig:MTF}
\end{figure}

For determining the MTF for $^{241}$Am $\alpha$-particle radiography, two Makrofol-KG\footnote{Markofol-KG is a mono-axially oriented and crystallized polycarbonate film with the chemical formula C$_{16}$H$_{14}$O$_{3}$ and a density of 1.2 g/cm$^3$.} foils of 10\,\textmu m thickness were attached to the backside electrode of the silicon sensor. The second foil was placed on top of the first one thus creating three regions of different penetration depths for the $\alpha$-particles before arriving at the detector. Both foils were placed so that their edges are at a small angle with respect to the $x$-axis of pixel matrix. A $^{241}$Am $\alpha$-source was placed at a distance of 10\,cm from the common electrode (backside) of Timepix3. The bias was set at 200\,V. Figure~\ref{fig:MTF} shows the energy of the $\alpha$-particles after the object, mapped out across the pixel matrix. The Kapton tape used for fixation of the set of foils is too thick to allow for $\alpha$-particle penetration, hence the existence of two white regions at both sides of the sensor. The other three regions from top to bottom depict the 20\,\textmu m, 10\,\textmu m foil and the area left uncovered, respectively. While for determination of the MTF the edge between the 10\,\textmu m Makrofol and the uncovered region are shown, the same results were found for the edge defined by the 10\,\textmu m and 20\,\textmu m regions (see Fig.~\ref{fig:annex_mtf}). The line spread function revealed a full-width at half maximum of $(21.7 \pm 2.4)$\,\textmu m, leading to a determined spatial frequency of $(20.6 \pm 1.2)$ line pairs per millimeter (lp/mm) at 10$\%$ MTF, twice the 9.09 lp/mm Nyquist frequency of a 55-\textmu m pixel pitch imaging device. The value found can be regarded as an upper bound of the intrinsic spatial resolution of Timepix3 considering the experimental design with a cone-beam geometry and scattering in the foils both degrading the edge response.
\begin{figure}[htbp]
\centering
\includegraphics[width=.89\textwidth, clip]{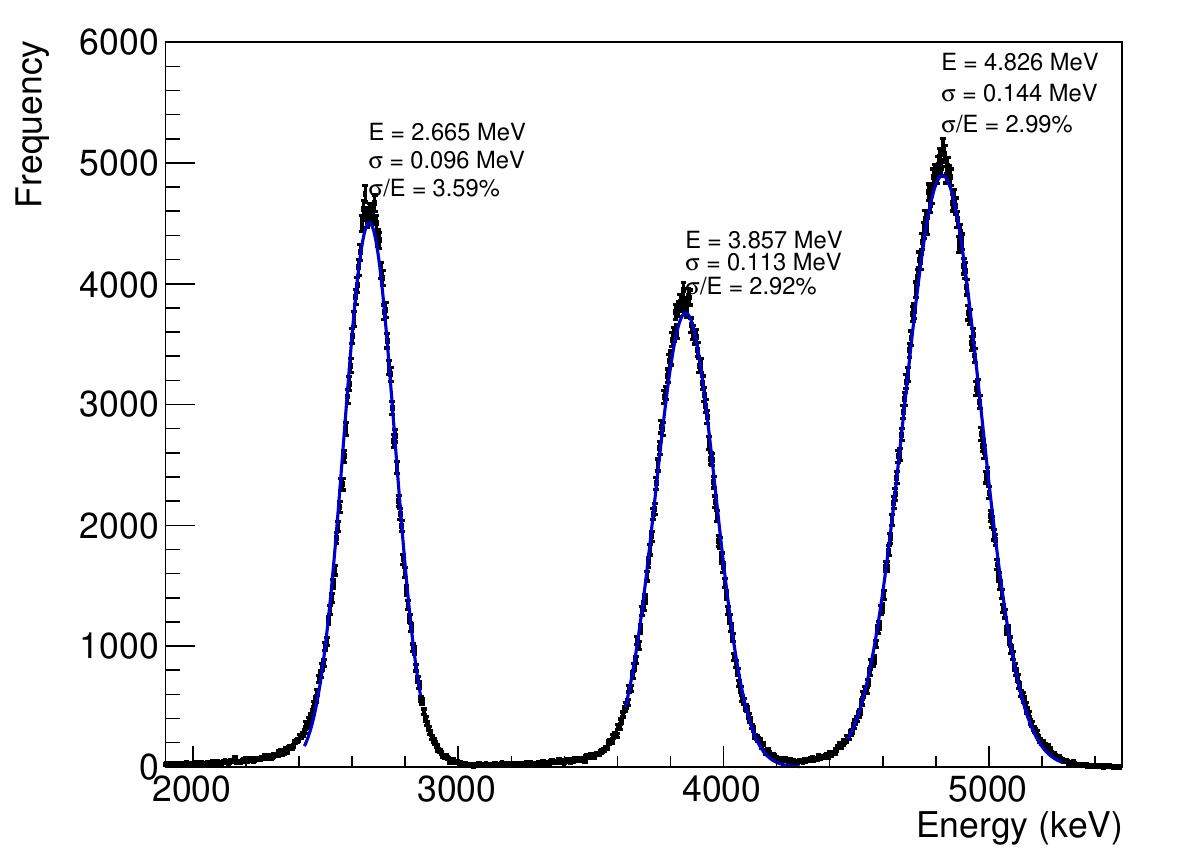}
\caption{$\alpha$-particle energy spectrum after passing through Makrofol foils. The peaks correspond to the different regions of different Mmakrofol foil thicknesses attached to the sensor backside as mapped out in Fig.~\ref{fig:MTF}.}
\label{fig:mylar_resolution}
\end{figure}

\subsubsection{Thickness resolution}
Figure~\ref{fig:mylar_resolution} shows the spectrum of $\alpha$ particle energies behind the Kapton foils. The peak of lowest energy represents the region covered with 20\,\textmu m while the peak of highest energy resembles the uncovered region. The resolution in the covered regions is reduced due to the scattering of the $\alpha$-particles in the Makrofol foils. Since the energy loss with sample thickness has linear behavior in the investigated energy range, we can access the achievable depth resolution $\sigma_{\rm depth}(E)$ as~\cite{tpx2adaptive}
\begin{equation}
    \sigma_{\rm depth}(E) = \frac{10\,{\rm \mu m}}{E(10\,{\rm \mu m})-E(20\,{\rm \mu m})} \times \sigma_{\rm E}(E).
\end{equation}

\noindent which gives values of $\sigma_{\rm depth}(4.9\,\textrm{MeV})$~=~1.33~\textmu m, $\sigma_{\rm depth}(3.9\,\textrm{MeV})$~=~1.16~\textmu m and $\sigma_{\rm depth}(2.7\,\textrm{MeV})$~=~0.8~\textmu m, at 0\,\textmu m, 10\,\textmu m and 20\,\textmu m sample thickness respectively.

\subsubsection{Real-world samples}
Post-calibration radiography measurements were performed in vacuum to demonstrate the reliability of the method. The above used 5.5\,MeV ${}^{241}$Am source of 10\,kBq was placed at a distance of 10\,cm from the backside (common electrode) of the sensor to which the samples were attached with a Kapton tape. The bias voltage was set at 200\,V. A liquid cooling loop was used to eliminate energy-temperature variations by keeping the sensor at $\sim$50 $^\circ$C, the temperature at which all calibration data were taken. For the image, each cluster event is assigned a physical pixel on the basis of its determined energy-weighted centroid. The pixel matrix is then filled with the average energy registered for each pixel. Where sufficient statistics is available, oversampling is performed by splitting the total events registered by one pixel into n$^2$ equally proportional sections. 

\begin{figure}[htbp]
\centering
%\includegraphics[width=.3\textwidth]{fly_photo_1.jpg} ${------ }$
%\qquad
%\includegraphics[width=.3\textwidth]{onion_photo.jpg}
\includegraphics[width=.49\textwidth]{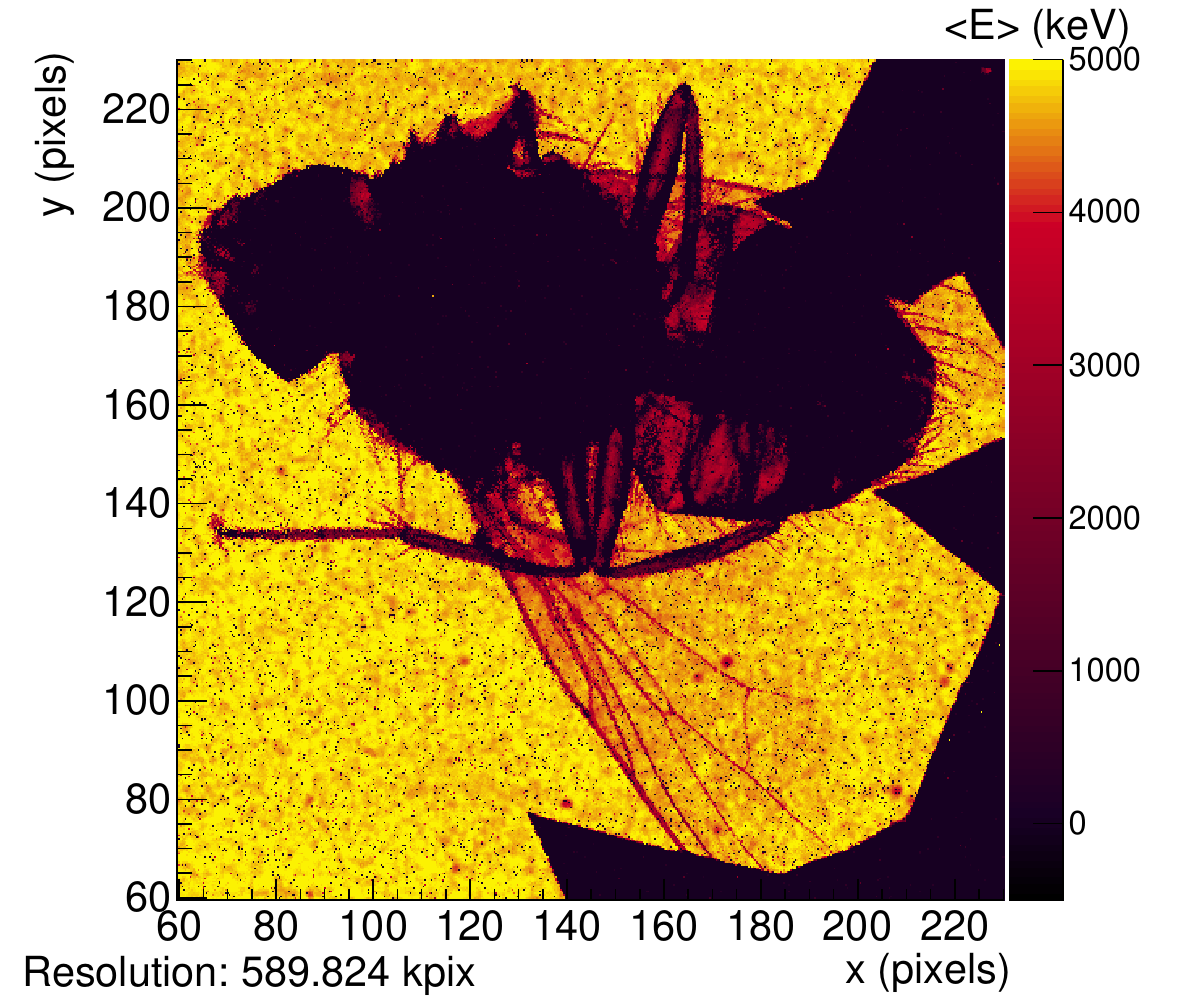}
\includegraphics[width=.49\textwidth]{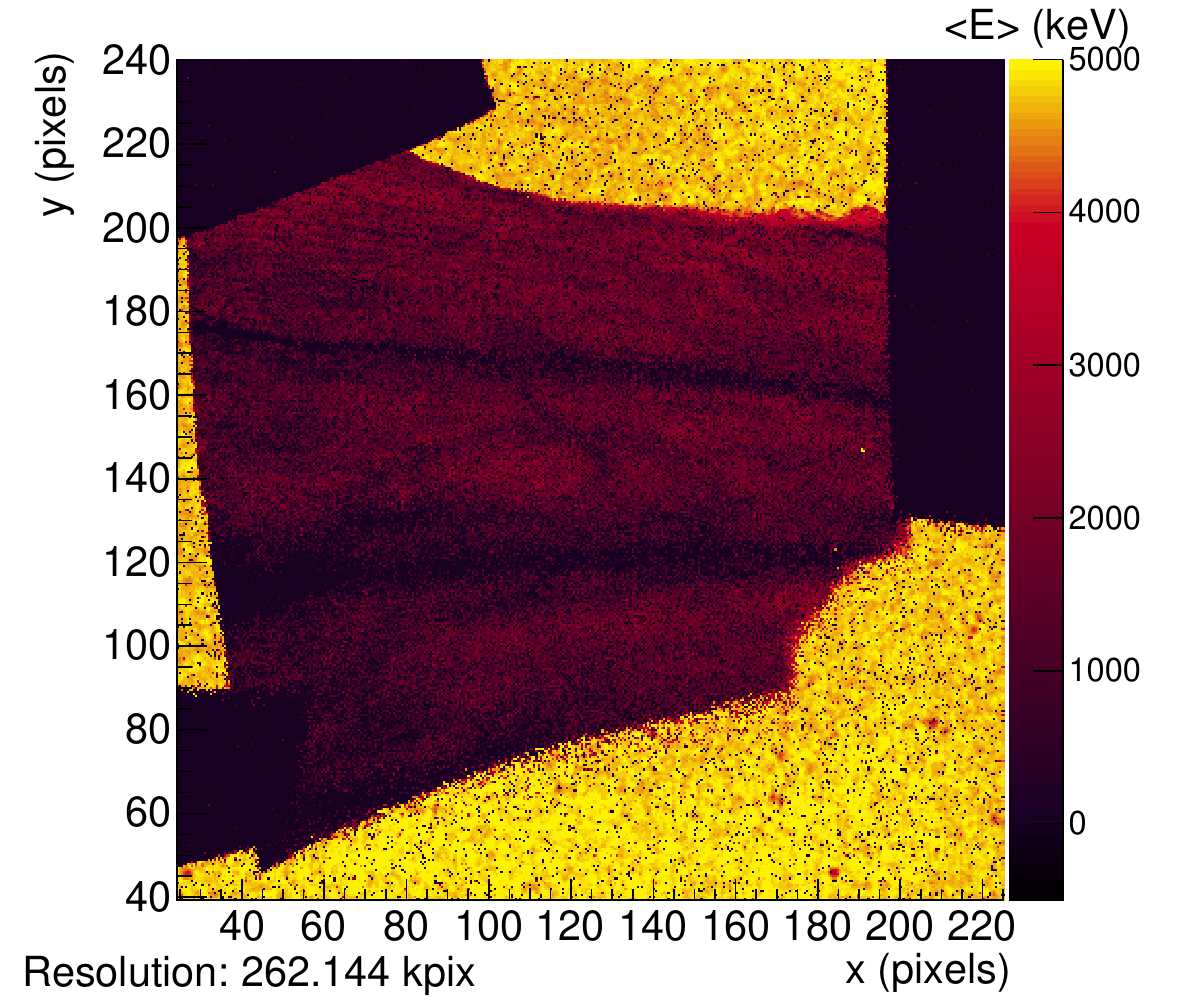}
\caption{$^{241}$Am alpha radiography images of a house fly (\emph{Musca domestica}, \textit{left}) and a portion of the skin of a red onion (\emph{Allium cepa}, \textit{right}), taken with a 500-\textmu m-thick silicon Timepix3 detector in vacuum. The bias voltage was set at 200\,V.}
\label{fig:radiography}
\end{figure}

\section{Conclusions}
The present work presented a study of the Timepix3 per-pixel response at high input charge. While the maximal distinguishable per-pixel energy deposition for Timepix and Timepix2 has previously been identified as 1.8\,MeV and 3.2\,MeV, respectively, Timepix3 fell well behind with its assumed value of 500 keV. Upon applying an energy-measurement correction technique, it was found that the reliable per-pixel energy range of Timepix3 can be increased to at least 1.1 MeV. After application of the correction, a relative energy resolution better than 2.5\% was found for stopped protons of up to 1.9 MeV and a resolution better than 3.1\% for $\alpha$-particles of 5.5\,MeV. Studies employing Timepix3 detectors with different silicon sensors and the use of different settings provided a consistent shape of the energy correction function, indicating a high level of reproducibility and the reliability of the proposed correction method. 

The thereby increased range of the per-pixel energy measurement significantly extends the versatility of Timepix3, improving its use in particle spectroscopy measurements for fundamental nuclear research and spatially-precise low-energy ion radiography measurements, making it a favored option in comparison to its predecessors, especially considering its low dead time continuous running mode and the improved time resolution.

The devices capabilities were evaluated for radiography with 5.5\,MeV $\alpha$-particles. The MTF of a slanted-edge sample indicates that subpixel resolution can be achieved while the sample-thickness variations could be resolved with a resolution of $\sim$1\,\textmu m.

\acknowledgments
The authors thank Ivan Wilhelm, Rudolf S\' ykora and Zden\u ek Kohout for their constant support during the ion measurements at the IEAP CTU in Prague Van-de-Graaff facility. B.B. and P.S. acknowledge funding from the Czech Science Foundation under Registration Number GM23-04869M.

\newpage

\appendix
\section{Results from the 300\,\textmu m thick sensor}

\begin{table}[htbp]
\centering
\caption{Properties of incident particles used in the calibration procedure for a Timepix3 detector with a 300-\textmu m silicon sensor. \label{tab:j}}
\smallskip
\begin{tabular}{c|c|c|cc}
\hline
\hline Particle & Energy at emission & Energy after RBS & Energy after dead layer \\
 & (keV) & (keV)& (keV) \\ 
\hline
%p & 500 & 490 & 405   \\
p & 600 & 588 & 511   \\
 & 700 & 685 & 615   \\
 & 800 & 784 & 718   \\
 & 1000 & 980 & 922   \\
 & 1500 & 1469 & 1425   \\
 & 1750 & 1714 & 1675   \\
 \hline
$\alpha$ & 5468 & - & 5281 \\
\hline
\end{tabular}
\end{table}

\begin{figure}[htbp]
\centering
\includegraphics[width=.49\textwidth]{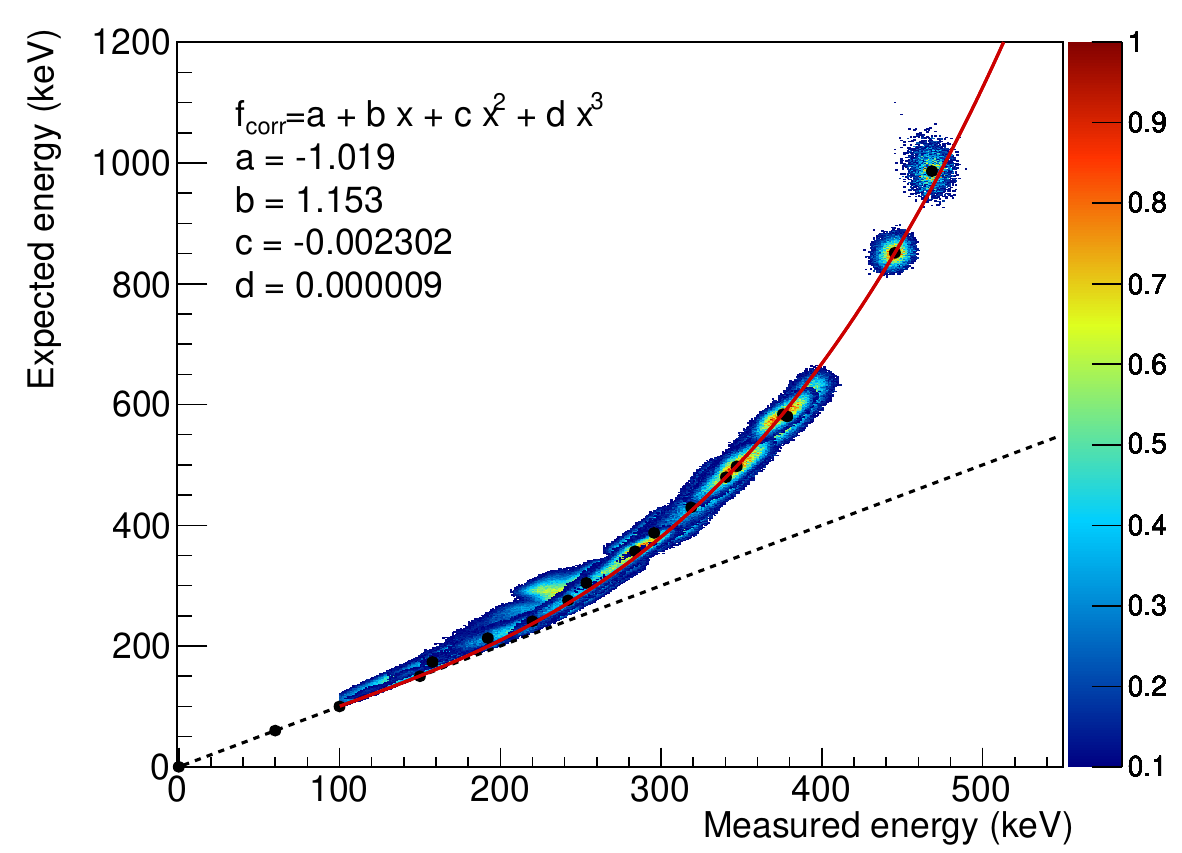}
\includegraphics[width=.49\textwidth]{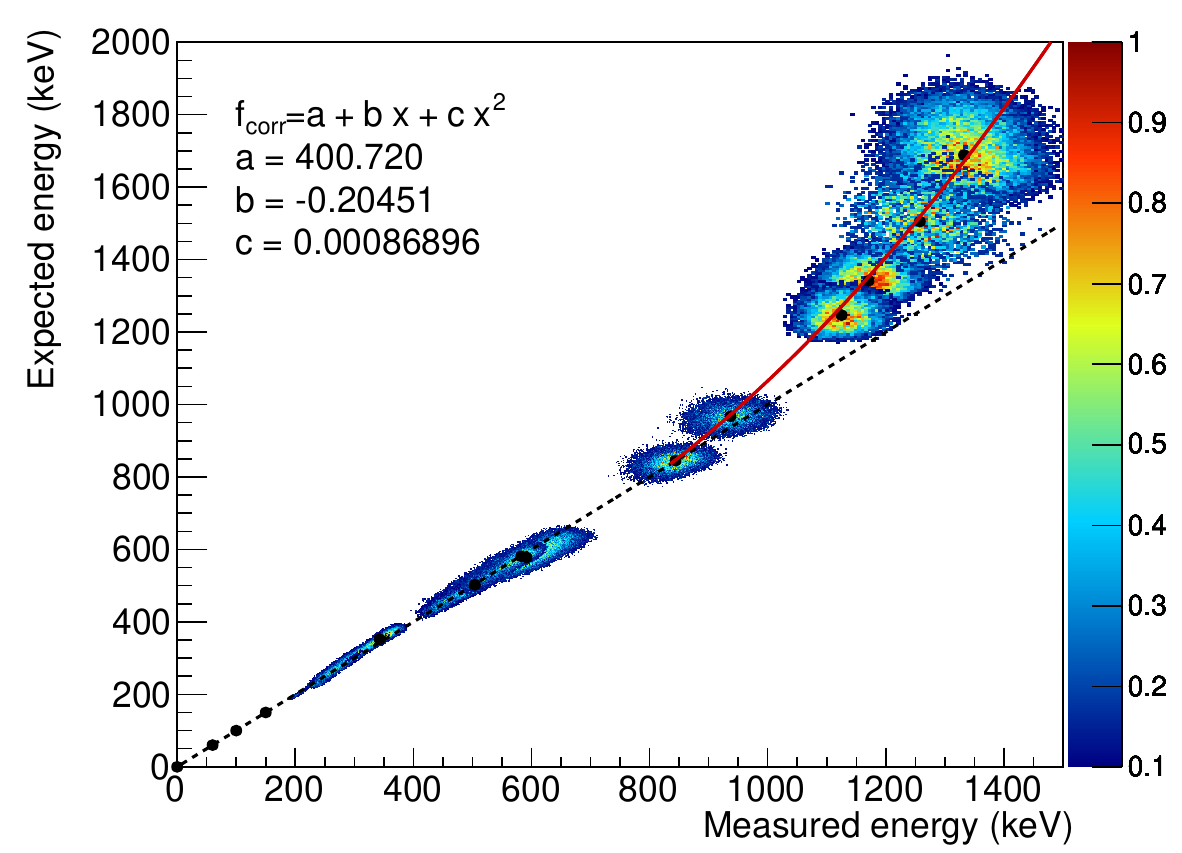}

\caption{Global calibration curves (red) for events registered by individual Timepix3 chip pixels connected to a 300-\,\textmu m silicon sensor. Due to reduced charge sharing compared to thicker sensors, expected per-pixel energies of 1.75-MeV protons (\textit{left}) are already much closer to the Timepix3 characteristic saturation, reaching values up to 1 MeV. The second iteration (\textit{right}) was called over $\alpha$-particle data measured at bias voltages ranging 100-150V.
\label{fig:iteration1_300}}
\end{figure}

\begin{figure}[htbp]
\centering
\includegraphics[width=.89\textwidth]{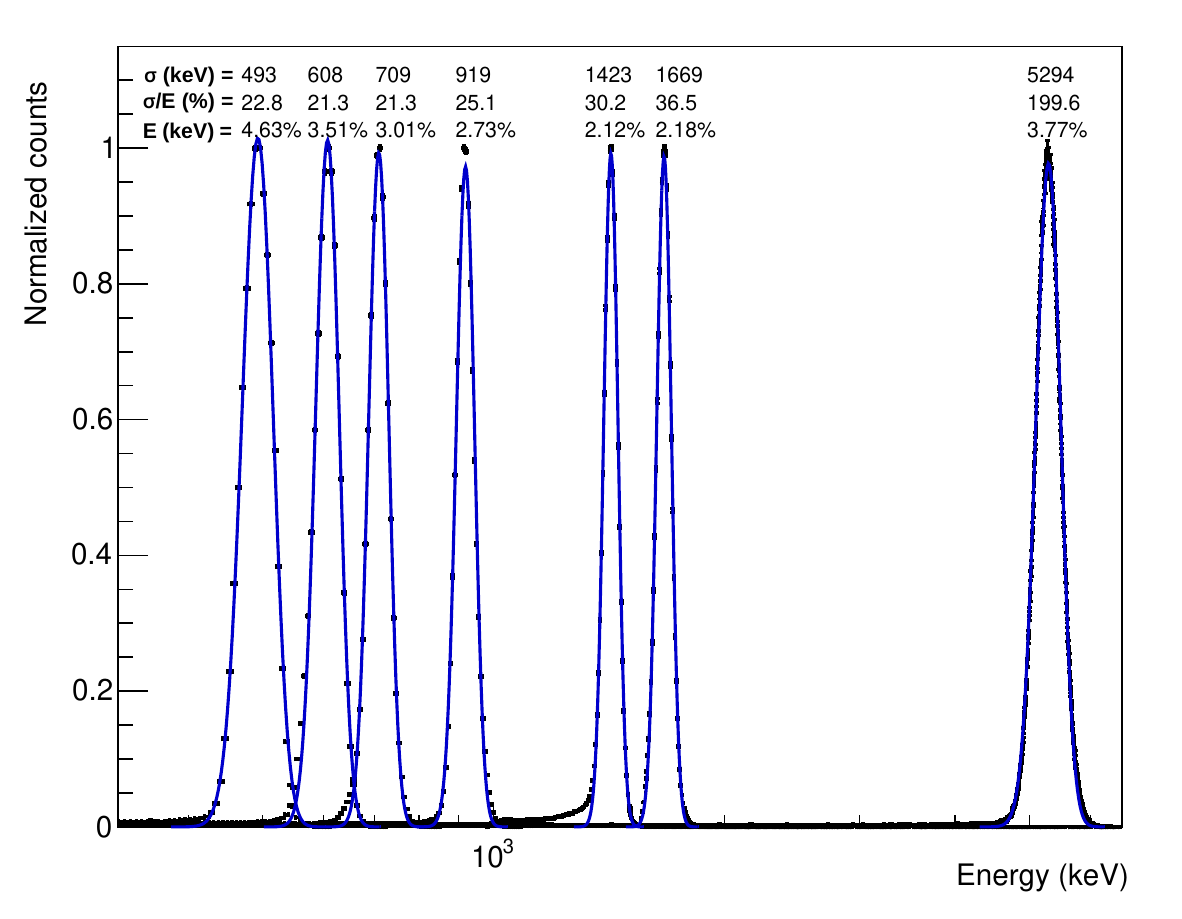}
\caption{Energy spectra of experimentally measured incident protons and alpha particles with a 300-\textmu m Si Timepix3 detector after applying the first iteration correction described in the present work. Intensities are scaled to the lowest height.
\label{fig:energy_calibration_300}}
\end{figure}
\newpage
\section{Slanted edge response at larger foil thickness}

\begin{figure}[htbp]
\centering
\includegraphics[width=.98\textwidth]{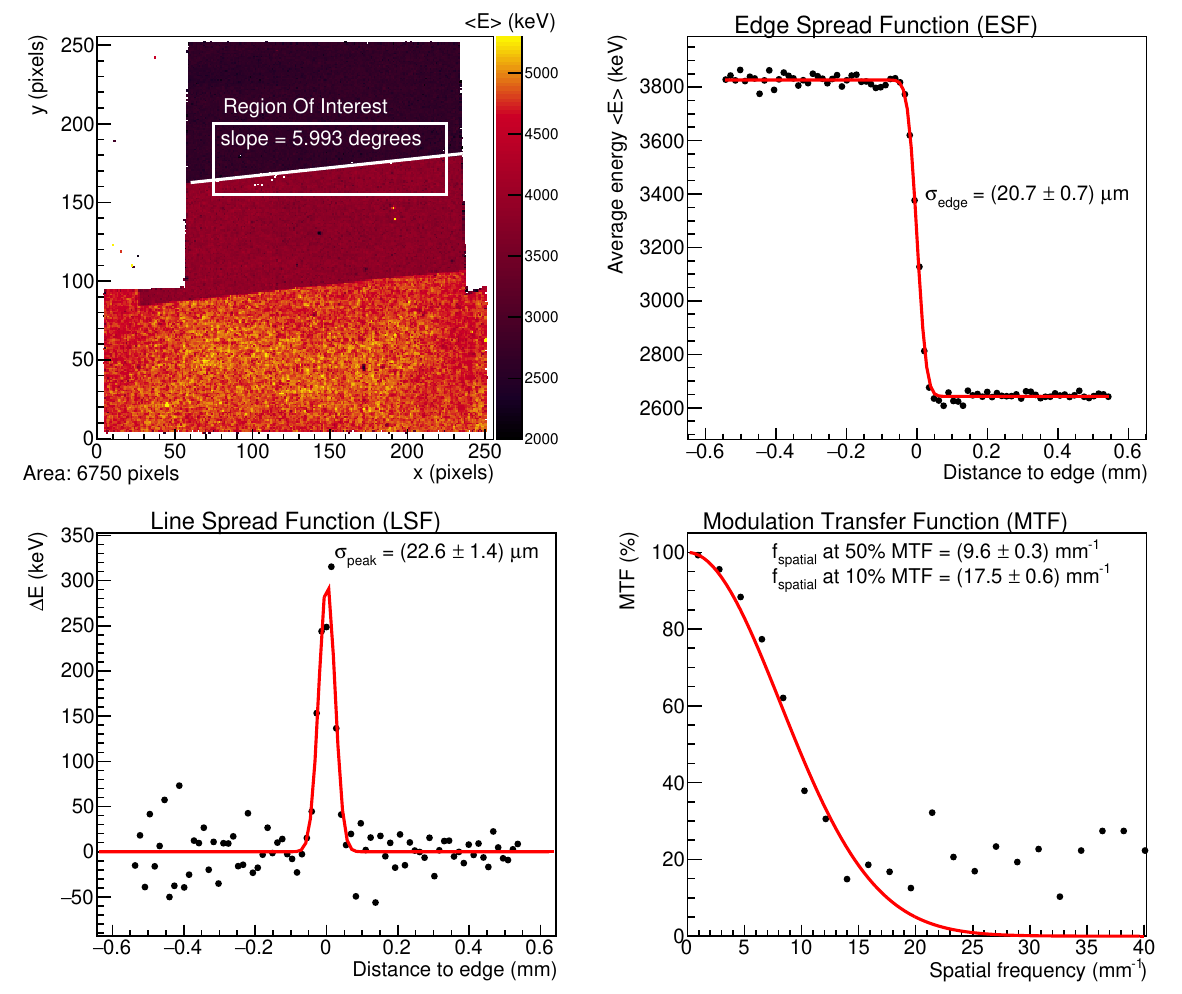}
\caption{Same as Fig.~\ref{fig:MTF}, but for the edge created by the 10\,\textmu m and 20\,\textmu m foils.}
\label{fig:annex_mtf}
\end{figure}

\end{document}